\documentclass[%
 reprint,
superscriptaddress,
 amsmath,amssymb,
 aps,
]{revtex4-2}

\usepackage{titlesec}
\usepackage{graphicx}
\usepackage{dcolumn}
\usepackage{bm}
\usepackage{hyperref}
\usepackage{xcolor}
\usepackage[T1]{fontenc}
\usepackage[utf8]{inputenc}
\usepackage{siunitx}
\usepackage{ragged2e} 
\usepackage[english]{babel}
\hypersetup{colorlinks,linkcolor={red!50!black},citecolor={blue!50!black},urlcolor={blue!80!black}}
\usepackage{pgf}
\usepackage{pgfplots}

\begin{document}

\preprint{APS/123-QED}

\title{Generation of Motional Squeezed States \\for Neutral Atoms in Optical Tweezers} 


\author{Vincent Lienhard}
\thanks{These authors contributed equally.}

\affiliation{Institute for Molecular Science, National Institutes of Natural Sciences, Okazaki 444-8585, Japan}
\affiliation{Nanomaterials and Nanotechnology Research Center (CINN-CSIC),
 Universidad de Oviedo (UO), Principado de Asturias, 33940 El Entrego, Spain}
 
\author{Romain Martin}
\thanks{These authors contributed equally.}

\affiliation{Institute for Molecular Science, National Institutes of Natural Sciences, Okazaki 444-8585, Japan}
\affiliation{Université Paris-Saclay, Institut d’Optique Graduate School, CNRS, Laboratoire Charles Fabry, 91127 Palaiseau Cedex, France}

\author{Yuki Torii Chew}
\affiliation{Institute for Molecular Science, National Institutes of Natural Sciences, Okazaki 444-8585, Japan}
\affiliation{Université Paris-Saclay, Institut d’Optique Graduate School, CNRS, Laboratoire Charles Fabry, 91127 Palaiseau Cedex, France}

\author{Takafumi Tomita}
\affiliation{Institute for Molecular Science, National Institutes of Natural Sciences, Okazaki 444-8585, Japan}
\affiliation{SOKENDAI (The Graduate University for Advanced Studies), Okazaki, Japan}
\author{Kenji Ohmori}
\affiliation{Institute for Molecular Science, National Institutes of Natural Sciences, Okazaki 444-8585, Japan}
\affiliation{SOKENDAI (The Graduate University for Advanced Studies), Okazaki, Japan}
\author{Sylvain de Léséleuc}
\email{sylvain@ims.ac.jp}
\affiliation{Institute for Molecular Science, National Institutes of Natural Sciences, Okazaki 444-8585, Japan}
\affiliation{RIKEN Center for Quantum Computing (RQC), RIKEN, Wako, Japan}

\date{\today}

\begin{abstract}

Optical tweezers are a powerful tool for the precise positioning of a variety of small objects, including single neutral atoms. Once trapped, atoms can be cooled to the motional ground state of the tweezers. For a more advanced control of their spatial wavefunction, we report here a simple method to squeeze their motion, and the protocol to measure the squeezing factor based on momentum spreading estimation. We explore the limitations set by the technical imperfections of the tweezers, as well as the more fundamental limit set by their anharmonicity, and finally demonstrate a squeezing of 5.8 dB.  
The implementation of motional squeezing allows to push back the limit set by the position quantum noise and thus to explore more extreme situations requiring atoms positioned with nanometric precision. 

\end{abstract}

\maketitle


Single neutral atoms trapped in optical tweezers~\cite{Ashkin2000,Schlosser2001,Kaufman2021} are widely used in numerous quantum technology applications, ranging from quantum sensing~\cite{Schaffner2024,Tomita2024}, quantum information networks with cavities~\cite{StamperKurn2023,Seubert2025} or waveguides~\cite{Luan2020,Menon2024}; to quantum simulation~\cite{Browaeys2020} and quantum computation~\cite{Bluvstein2024,Radnaev2024, Reichardt2024}. Part of the experimental effort of the community has been dedicated to reducing the residual motion of the atoms in the tweezers, which can be seen in good approximation as a harmonic oscillator, for a more accurate position control and higher fidelity of quantum operations. This is achieved by cooling the single atom to the motional ground-state by Raman sideband cooling, as first reported in Refs.~\cite{Kaufman2012,Thompson2013}. The atom is then described by a wavefunction with quantum uncertainties along the position $x$ or momentum $p_x$ quadratures saturating the Heisenberg inequality as $\Delta x^{\mathrm{gs}} = \sqrt{\hbar/\left(2m\omega_r\right)}$ and $\Delta p_x^{\mathrm{gs}}=\sqrt{\hbar m\omega_r/2} $, where $\omega_r$ is the angular frequency of the oscillator and $m$ the mass of the atom. For $^{87}\mathrm{Rb}$ atoms and typical trapping frequency of $\omega = 2\pi\times 100\,\mathrm{kHz}$, one obtains $\Delta x^{\mathrm{gs}} = 24\,\mathrm{nm}$ and $\Delta p_x^{\mathrm{gs}}=m\cdot 0.015\,\mathrm{ms}^{-1}$, which can be compared to a typical thermal uncertainty $\Delta x^{\mathrm{th}} = \sqrt{\mathrm{k}_\mathrm{B} T/\left(m\omega_r^2\right)} \approx 110\,\mathrm{nm}$ for a temperature $T= 50 \, \mu$K, or to the typical size of the tweezers, the waist, on the order of $0.5-1\,\mu$m.


A natural way to increase the positioning precision, beyond the standard quantum limit described above, consists in squeezing the state: decreasing the uncertainty along one quadrature at the expense of increasing it on the other. We then define the squeezing factor $\beta$ as $\Delta x^{\mathrm{sq}} = \Delta x^{\mathrm{gs}} / \beta$ and $\Delta p_x^{\mathrm{sq}} = \Delta p_x^{\mathrm{gs}} \times \beta$.
Squeezing was first developed in the context of quantum optics~\cite{Caves1981,Xiao1987,Grangier1987}, for metrological purposes, and is now a common tool experimentally used on a large diversity of quantum fields and platforms: motion of trapped ions~\cite{Cirac1993,Meekhof1996}, of atoms in optical lattices~\cite{Morinaga1999}, and even of levitating nanoparticles~\cite{Rossi2024}; or electromagnetic waves in superconducting circuits~\cite{Castellanos2008,Joshi2017} and in gravitational wave detectors~\cite{Abadie2011,Acernese2019}. As a few examples, displacement sensing through quantum amplification has been achieved by squeezing trapped ions by up to $20\,\mathrm{dB}$ ($\beta = 10$)~\cite{Burd2019}, a $8\,\mathrm{dB}$ squeezed microwave field was produced in a superconducting  cavity~\cite{Dassonneville2021}, and $8.5\,\mathrm{dB}$ squeezed light has been injected in gravitational wave interferometers~\cite{VIRGO2023}. 

In this work, we report a simple method to generate position-squeezed states for a collection of atoms individually trapped in optical tweezers. We first describe the generation and probing of squeezed states. We then show that optimizing the shape of the tweezers by holography~\cite{Chew2024} allows to push the achievable squeezing level in an array, by compensating the tweezers radial anisotropy and minimizing the shape inhomogeneity over the ensemble of traps. 
Finally, we discuss the limit set to the squeezing factor by the inherent anharmonicity of optical tweezers. 

\begin{figure}[t!]\centering
\includegraphics[width=.98\columnwidth]{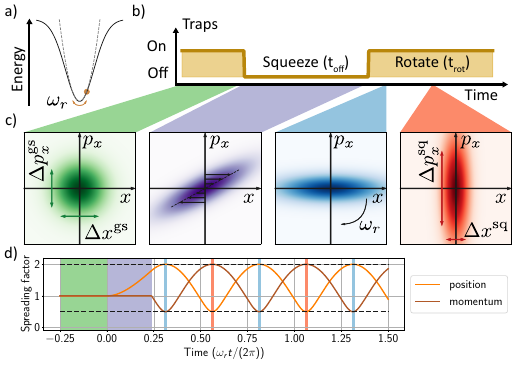}
\caption{ Generation of squeezed states. a) Optical tweezer potential (solid line), and its harmonic approximation (dashed line). b) Experimental sequence and c) sketches of the Wigner quasi-probability distribution of the atomic state in the $x$-$p_x$ phase space through the sequence. d) Theoretical standard deviations of $x$ and $p_x$, normalized to the motional ground-state fluctuations, as a function of time. The atom is released from a harmonic trap during a time $t_\mathrm{off}$, set here to $\omega_r t_\mathrm{off} \approx 1.5$ (giving $\beta = 2$, see Eq.~\ref{eq:eq3}).}
\label{fig:fig1}
\end{figure}

\paragraph*{Experimental platform} Our set up, presented in \cite{Chew2022,Chew2024}, is based on an array of $^{87}$Rb atoms individually trapped in holographic optical tweezers. These microtraps are created by tightly focusing a $852\,\mathrm{nm}$ laser beam via an objective of numerical aperture NA=0.75. The geometry of the optical tweezers array is controlled via a Spatial Light Modulator (SLM), impinging a versatile phase pattern (hologram) on the trap beam. Here, the chosen configuration is a $6\times 12$ array with typical distance of 5~$\mu$m between atoms. The atoms are loaded from a magneto-optical trap (MOT), at a temperature of $50\,\mu\mathrm{K}$, into the optical tweezers (tweezers depth $U_0/\mathrm{k}_\mathrm{B} = 0.5\,\mathrm{mK}$), thus occupying the bottom of the trap (see Fig.~\ref{fig:fig1}a). In a first approximation, we write the trapping potential as harmonic and radially symmetric:
\begin{equation}
\label{eq:eq1}
U\left(x,y,z\right) = -U_0 + \frac{1}{2}m\omega_r^2\left(x^2+y^2\right)+\frac{1}{2}m\omega_z^2 z^2,
\end{equation}
where the radial (motion along $x$ and $y$) and the longitudinal (motion along $z$) trapping angular frequencies are given by:
\begin{equation}
\label{eq:eq2}
\omega_r = \sqrt{\frac{4 U_0}{mw_0^2}},\quad \omega_z = \sqrt{\frac{2 U_0}{mz_\mathrm{R}^2}},
\end{equation}
with $w_0$ and $z_\mathrm{R}$ respectively the waist and the Rayleigh length of the tweezers. The typical trapping frequencies measured on our set-up are $\omega_r \approx 2\pi \times 105\,\mathrm{kHz}$ and $\omega_z \approx 2\pi \times 25\,\mathrm{kHz}$. Atoms loaded from the MOT are initially occupying high vibrational levels $\langle n_r \rangle \sim 10$.

\begin{figure}[t!]\centering
\includegraphics[width=.98\columnwidth]{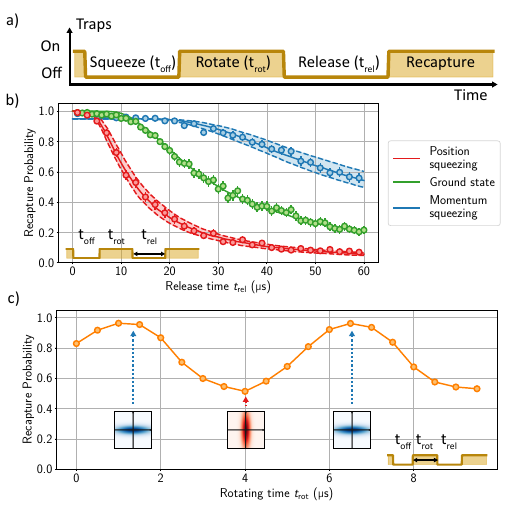}
\caption{Probing the squeezed states. a) Experimental sequence for generating a squeeze state, followed by a release-and-recapture probing. b) Release-and-recapture results for three different motional states. The two squeezed states were prepared for $t_\mathrm{off} = 2.0\,\mu\mathrm{s}$, and a trapping frequency $\omega_r \approx 2\pi \times 110\,\mathrm{kHz}$. Solid lines are simulations, the dashed lines and colored area indicate the error bar on the fitted $\beta$ (see main text). c) Oscillation of the squeezed state in the trap, performed for a slightly different trapping frequency $\omega_r \approx 2\pi \times 95\,\mathrm{kHz}$ ($t_\mathrm{off} = 2.0\,\mu\mathrm{s}$ and $t_\mathrm{rel} = 20\,\mu\mathrm{s}$). Data in b (c) are obtained by averaging the signals of 72 (50) tweezers over 15 (30) experimental runs. Error bars, which are almost always smaller than the data symbol, represent the standard error on the mean.}
\label{fig:fig2}

\end{figure}

\paragraph*{Squeezing protocol} First, the atoms are cooled down to the motional ground state via Raman sideband cooling~\cite{Kaufman2012,Thompson2013}. We observe a preparation efficiency of the radial motional ground state of more than $90\%$, more precisely $\left\langle n_{x,y} \right\rangle \approx 0.05$ and $\left\langle n_z \right\rangle \approx 0.3$~\cite{Chew2022}. The squeezing protocol is then as follows (see Fig~\ref{fig:fig1}b): we switch off the traps for a duration $t_\mathrm{off}$, letting the atom wavefunction expands, and switch the traps on again to make the atoms wobble inside them during a time $t_\mathrm{rot}$. The phase-space representations of the motional state of the atom along one of the two radial directions (say $x$) is displayed on Fig~\ref{fig:fig1}c and illustrate how the state is squeezed. At first, it is a circular distribution with quantum uncertainties at the standard quantum limit (green). Then, during the free-fly step (purple), the distribution is sheared along the horizontal axis, as points higher on the $p_x$ axis fly faster along the $x$ axis. Third, the traps are switched back on, making the distribution rotates with angular frequency $\omega_r$ between a momentum-squeezed state (blue) and a position-squeezed state (red). Consequently, the uncertainties along $x$ and $p_x$, normalized by $\Delta x^{\mathrm{gs}}$ and $\Delta p_x^{\mathrm{gs}}$, show out-of-phase oscillations between a maximal value $\beta$ and a minimal value $1/\beta$ (black horizontal dashed lines in Fig~\ref{fig:fig1}d), at twice the trapping frequency. 

The simple method described here allows for the preparation of a motional squeezed state. It was first demonstrated for neutral atoms in optical lattices in Ref.~\cite{Morinaga1999}, where the expression of $\beta$ was derived:
\begin{equation}
\label{eq:eq3}
\beta^2 = 1 + \frac{\left(\omega_r t_\mathrm{off}\right)^2}{2} + \omega_r t_\mathrm{off} \sqrt{1+\frac{\left(\omega_r t_\mathrm{off}\right)^2}{4}}, 
\end{equation}
with the long-time limit, $\beta \approx \omega_r t_\mathrm{off}$, being simply the ratio of the spatial expansion of the distribution due to its speed $\Delta p_x^{\mathrm{gs}} t_\mathrm{off}/m$ to its initial spread $\Delta x^{\mathrm{gs}}$. 



Switching off the traps is equivalent to adding a squeezing term $-\hat{x}^2 \propto \left({\hat{a}} + {\hat{a}}^{\dag }\right)^2 $, where $\hat{a}$ (${\hat{a}}^{\dag}$) are the canonical lowering (raising) operators. Squeezing can also be obtained by modulating the trap depth (instead of switching it off fully) at twice the trapping frequency~\cite{Burd2019}, or by performing a two-photon Raman drive~\cite{Meekhof1996}.
The advantage of the on/off approach chosen here is its simplicity: in the case of optical tweezers, it is possible to switch off in $10\, \mathrm{ns}$ (three orders of magnitude faster than the trap period). It also squeezes much faster, thus minimizing the effect of imperfection from the traps (see later).

\paragraph*{Observation of squeezed states} Squeezed states can be characterized either in the $n$ basis, or in the $x$-$p_x$ basis. For the first case, the populations and relative phases of the different vibrational levels are measured, as done in Ref.~\cite{Meekhof1996} by probing a motion-sensitive electronic transition. 
The other way consists in reconstructing the phase-space distribution. A direct method applicable for the optical tweezers platform was proposed in Ref.~\cite{Winkelmann2022}, while the more standard tomography approach was realized using time-of-flight measurements~\cite{Brown2023}

Here, we chose to measure only the momentum distribution uncertainty, and by observing its oscillation as the distribution rotates (see Fig.~\ref{fig:fig1}d), we will infer the squeezing factor for the position distribution~\cite{Asteria2021}. The rms (root-mean-square) velocity of the atoms is estimated with a standard release-and-recapture experiment and classical Monte-Carlo simulation~\cite{Tuchendler2008}. The experimental sequence is shown in Fig.~\ref{fig:fig2}a: after having squeezed and rotated the state, we release the atom again for a varying time $t_\mathrm{rel}$ before trying to recapture it. 
Figure~\ref{fig:fig2}b shows the measured recapture probability as a function of $t_\mathrm{rel}$ for three different states: the motional ground state (green), and two squeezed states obtained for $t_\mathrm{off} = 2.0\,\mu\mathrm{s}$ corresponding to an expected squeezing factor $\beta = 1.9$. The squeezed state with minimal momentum uncertainty is obtained for $t_\mathrm{rot} = 1.4\,\mu\mathrm{s}$ (blue), and the one with minimal position uncertainty for $t_\mathrm{rot} = 3.6\,\mu\mathrm{s}$ (red), i.e. after an additional $\pi/2$ rotation in phase space. The data for the ground state (green) serve as a reference for fixing the parameters of the Monte-Carlo simulation (solid line). 
For the blue data, the slower decay indicates a momentum squeezing, which is well captured by the simulation using a reduced momentum uncertainty by a best fitting factor $\beta = 1.96$. Whereas for the red data, the decay is faster and matches with an increased momentum spread by the same factor. The best fitting factor, and the associated error bar, are estimated via the $\chi^2$-method, and gives $\beta = 1.96 \pm 0.13$, in agreement with the expected value 1.9.
So far, we have reported the generation of a state whose momentum spreading is reduced and then increased by a factor 1.96. To further infer that the position uncertainty is reduced by this same factor, we need the distribution to rotate unperturbed, which is true only for an ideal harmonic potential. We can check this experimentally by observing if the momentum uncertainty periodically recovers its minimal value as the distribution rotates with increasing $t_{\rm rot}$. For example, we see that it barely decreases over half a trapping period in Fig.~\ref{fig:fig2}c. 


\begin{figure}[t!]\centering
\includegraphics[width=.98\columnwidth]{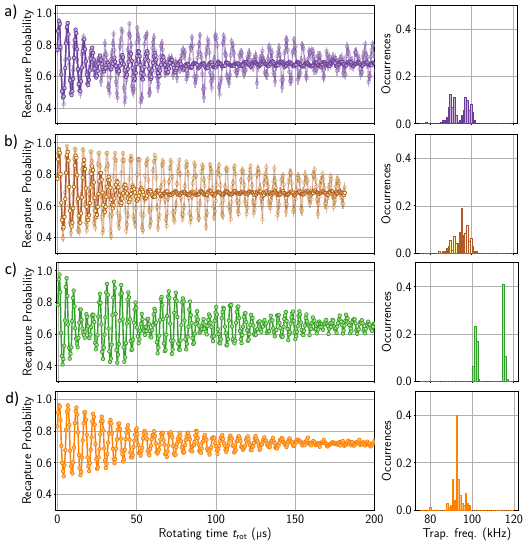}
\caption{Oscillation of squeezed states averaged over an array of optical tweezers. The displayed data correspond to the experimental parameters  $t_\mathrm{off} = 2.0\,\mu\mathrm{s}$ and $t_\mathrm{rel} = 20\,\mu\mathrm{s}$, while varying the rotating time $t_\mathrm{rot}$. The subfigures show the recapture probability for different corrections : a) no corrections, b) anisotropy compensation, c) inhomogeneity compensation and d) both corrections. The trapping frequencies, extracted from Fourier transforms of the individual atomic signals (shown in the lighter curve in a) and b)), are plotted in the right panels. For c) and d), we used an atomic array with 50 instead of 72 atoms, averaged over 30 iterations compared to 200 iterations for a) and b).}
\label{fig:fig3}
\end{figure}

\paragraph*{Collective oscillation of the squeezed states} 

Observing the squeezed state evolution for longer times, up to 20 trapping periods ($t_{\rm rot} \simeq 200 \, \mu$s), reveals striking features. 
In Fig.~\ref{fig:fig3}a, the oscillation displays a beatnote and an overall damping. This can be explained from the measured distribution of trapping frequencies over the array of tweezers (shown on the right panel):
the distribution is bimodal (inducing beating), and with a finite spread (inducing damping). The bimodality comes from a slight radial anisotropy (10-20\,\%) between the $x$ and $y$ axis, which naturally occurs at high NA~\cite{Wolf1959,Richards1959}.
The inhomogeneities over the array come from imperfections of the hologram. We also display typical single-atom signals, where the damping is much reduced confirming that the damping of the averaged signal comes from the inhomogeneities.


Counteracting on the SLM phase pattern allows us to compensate for the anisotropy and the shape inhomogeneities, as we recently demonstrated~\cite{Chew2024}. After anisotropy compensation, the resulting oscillation (Fig.~\ref{fig:fig3}b), shows no more beatnote and only damping (again, much faster than for a single atom), indicating that the two radial modes have now the same trapping frequency, which is confirmed by the distribution of individually measured trapping frequencies (see side panel). Alternatively, if we modify the phase pattern to decrease the frequency spread over the array, decreasing it by one order of magnitude to $\Delta \omega / \omega < 0.01$, we now observe a much reduced damping and a clearer beatnote (Fig.~\ref{fig:fig3}c). 
Finally, when applying both anisotropy and inhomogeneity corrections (Fig.~\ref{fig:fig3}d), we minimize the degradation of radial squeezing over time.

If position-squeezed state along only one of the two radial direction is needed~\cite{Chew2022}, then correcting the anisotropy is not required and we can rather optimize the inhomogeneity as in c). However, since our release-and-recapture measurement protocol does not distinguish between the $x$ and $y$ velocities, we rather chose to synchronize both radial modes, at the cost of 
a slightly worse homogeneity, which however will not be limiting anymore as we shall now explore more quantitatively. 

\begin{figure}[t!]\centering
\includegraphics[width=.98\columnwidth]{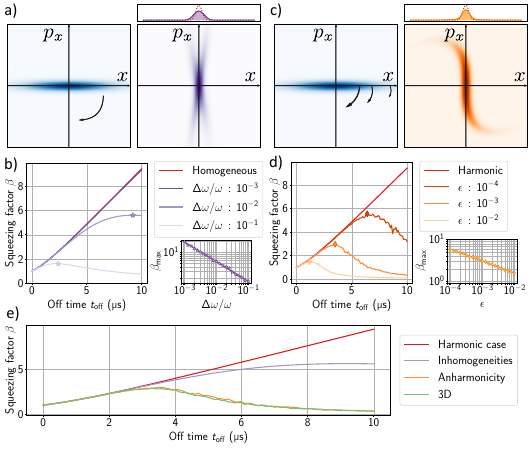}
\caption{Squeezing performance. a) Sketch of the effect of the trapping frequency inhomogeneity on the squeezing factor. Here we show the rotation of two motional states for a trapping frequency difference $\Delta \omega / \omega = 0.2$, and the corresponding projection on the $x$-axis (the red dashed line represents the ideal squeezed state). b) Squeezing factor as a function of $t_\mathrm{off}$, for different trapping frequency inhomogeneities. The inset shows the maximal squeezing factor as a function of $\Delta \omega / \omega$. c,d) Same as a,b) for the study of the anharmonicity. The anharmonicity is $\epsilon = 10^{-2}$ in c). e) Squeezing factor as a function of $t_\mathrm{off}$, in an ideal 1D harmonic potential (red), in the presence of our remaining inhomogeneities $\Delta \omega / \omega = 10^{-2}$ (purple), for the non-approximated 1D Gaussian potential (orange) and finally for the actual 3D potential (green).}
\label{fig:fig4}
\end{figure}

\paragraph*{Squeezing limitations} In this final part, we calculate how much squeezing in position can be achieved given the imperfect rotation of the squeezed phase-space distribution while the atoms oscillate in the traps. As just observed, a first limitation comes from different frequencies for the radial mode. In Fig.~\ref{fig:fig4}a, we consider two radial modes (for example, from two different traps), whose relative frequency difference is $\Delta \omega / \omega = 0.2$, initially squeezed in momentum by a factor $\beta = 3$. Due to their different angular speed, the two rotating distributions will not be aligned to the $p_x$ axis at the same time, leading to a broadening of the position distribution averaged over the two traps as compared to an ideal position-squeezed state (see inset of Fig.~\ref{fig:fig4}a showing the projection of the phase-space distribution on the $x$ axis). For a given value of $\Delta \omega / \omega$, this sets a maximal squeezing factor $\beta_\mathrm{max}$ that can be achieved over the atomic array.  
A scaling law can be obtained by equating the $x$-projection of the imperfectly-rotated distribution $\beta_\mathrm{max} \mathrm{sin}\left(\Delta \omega t_\mathrm{rot}\right)$ to the ideal squeezing value $1/\beta_\mathrm{max}$. Using $t_\mathrm{rot} \propto 1/\omega$, we finally get: $\beta_\mathrm{max} \propto \left(\Delta \omega /\omega\right)^{-1/2}$.

We then perform numerical calculations by evolving squeezed states in 1D harmonic potentials with trapping frequencies distributed following a Gaussian distribution with an rms of $\Delta \omega$ (Figure~\ref{fig:fig4}b). We extract a maximal squeezing factor (indicated by the star symbols) that indeed scales as $\beta_\mathrm{max} \propto \left(\Delta \omega /\omega\right)^{-0.53}$.
For the uncorrected hologram, with $\Delta \omega /\omega \sim 0.1$, the squeezing factor would be limited to $\beta_{\rm max} = 1.6$. After optimization of the hologram, which decreases the inhomogeneity by an order of magnitude, we could hope to reach a squeezing of more than 5 (14 dB). 

We now consider a second, more inherent, imperfection of the tweezers: its anharmonicity. 
As seen in Fig.~\ref{fig:fig1}a, the harmonic approximation of the tweezers potential by a parabola breaks down when the atom explores a too large region. This results in a distortion of the phase-space distribution, shown in Fig.~\ref{fig:fig4}c, as the atom evolves more slowly in the sub-harmonic region.
This anharmonicity can be captured by expanding a 1D Gaussian potential to the fourth order in $x$:
\begin{equation}
\label{eq:eq4}
U\left(x\right) = -U_0 + \frac{1}{2}m \omega_\mathrm{r}^2 x^2 \left(1-\epsilon\left( \frac{x}{\Delta x ^\mathrm{gs}}\right)^2\right),
\end{equation}
where $\epsilon = \hbar \omega_\mathrm{r} / \left(8 U_0\right) \approx 10^{-3}$ for our tweezers parameters. 
We can then again derive a scaling law for the maximal squeezing factor $\beta_\mathrm{max}$ achievable for a given anharmonicity parameter $\epsilon$. 
We estimate the difference in angular frequency for the center of the distribution as compared to its "tip" (at $x = \beta$) as $\Delta \omega = \omega \left(1-\sqrt{1-\epsilon \beta^2}\right)$, thus leading to $\beta_\mathrm{max}\propto \epsilon^{-1/4}$.
Numerical results, shown in Fig.~\ref{fig:fig4}d, are obtained by evolving a squeezed state in 1D Gaussian potential with fixed trapping frequency $\omega$ but varying anisotropy parameter $\epsilon$ (obtained by adjusting the trap depth $U_0$ and waist $w$ accordingly). We obtain $\beta_{\rm max} = 3$ at $\epsilon = 10^{-3}$ and a power law of $\beta_\mathrm{max}\propto \epsilon^{-0.28}$.

Finally, we compare these different limitations for our tweezers parameters in Fig.~\ref{fig:fig4}e, illustrating again that the anharmonicity becomes the limit once the traps shape are optimized by holography. 
We have also performed a simulation with the full 3D potential of the tweezers, showing no difference with the 1D case, justifying \emph{a posteriori} why we have not tackled it in our previous discussions. There is also squeezing along the longitudinal direction, but it is almost negligible ($\beta_z = 1.1$ at $t_{\rm off} = 2 \mu$s) given the much smaller trapping frequency $\omega_z$ in that direction, and would give no observable signature on the release-and-recapture experiment which is mostly sensitive to the radial motion. 

\paragraph*{Discussions} 
Having analyzed the limitations set by the trap shape, we now conclude on the squeezing level demonstrated in Fig.~\ref{fig:fig2}. For the chosen experimental value $t_\mathrm{off} = 2\,\mu\mathrm{s}$, the numerical results of Fig.~\ref{fig:fig4}e demonstrates that the anharmonicity is not yet a limit. In addition, the results in Fig.~\ref{fig:fig2}b were obtained with an optimized hologram, so that trap anisotropy and inhomogeneities does not play a role. This allows us to claim that the squeezed distribution oscillates unperturbed on the timescale of interest, and thus to conclude that we generated a position-squeezed state with a quantum uncertainty of $\Delta x^{\rm sq} = 13$~nm, reduced below the standard quantum limit by 5.8 dB ($\beta = 1.96$).

A more straightforward approach to reduce the position uncertainty of the wavefunction could have been simply to increase the trapping frequency. But this would not be a scalable strategy: as $\Delta x^{\mathrm{gs}} \propto U_0^{-1/4}$, reducing the position spreading by a similar factor of 2 would have required an increase of the power by 16. 
We also note that deeper tweezers reduces the anharmonicity parameter ($\epsilon \propto U_0^{-1/2}$), but will also not be a scalable solution to push further the achievable squeezing given the very disadvantageous scaling of $\beta_{\max} \propto \epsilon^{-1/4}$.  

Finally, we envision several applications for these position-squeezed states. Given that they are not static, we are looking at applications on timescale of a few hundred nanoseconds at most.
Using them in combination of strong Rydberg interactions, which can reach a strength of several hundreds of MHz~\cite{Chew2022,Bharti2024}, is a first direction. 
They could allow the engineering of more complex Hamiltonians or the use of motional degrees of freedom in quantum information processing via spin-motion coupling~\cite{Gambetta2020,Mazza2020,Magoni2023,Mehaignerie2023,Zhang2024,Nill2025,Burd2024,Shaw2024}. 
We could take advantage of the squeezed states to reduce the decoherence of spin-exchange oscillation~\cite{Chew2022,Emperauger2025}. We could also use quantum amplification based on squeezing to measure the momentum kick between Rydberg atoms~\cite{Burd2019}. Quantum amplification requires to also unsqueeze the state, which can be simply realized on our set-up by repeating the squeezing protocol again once the distribution is in the mirror-image with respect to the $p$-axis after rotation~\cite{Kim2025}.
A second direction is to combine the squeezing technique with flying atoms~\cite{Hwang2023} to prepare ordered arrays well within the sub-wavelength regime where fascinating collective light-matter phenomena emerges~\cite{Yelin2022,Masson2022}. Finally, one could also use the anharmonicity for the generation of non-classical states~\cite{Kendell2023,Grochowski2024,Taniguchi2025}.

\begin{acknowledgments}
This work was supported by MEXT Quantum Leap Flagship Program
(MEXT Q-LEAP) JPMXS0118069021 and JST Moonshot R\&D Program Grant Number JPMJMS2269. VL acknowledges support by the European Commission – NextGenerationEU, through Momentum CSIC Programme: Develop Your Digital Talent.
\end{acknowledgments}


\bibliography{paper}

\end{document}